\documentclass{kapproc} 

\usepackage{procps} 

\usepackage[dvips]{graphicx}

\upperandlowercase

\kluwerbib 

\newcommand{\tal}{\it et al. \rm}

\begin{document}



\articletitle[]{Halo properties and secular \\evolution in
  barred galaxies}


\author{E. Athanassoula}
\affil{Observatoire de Marseille Provence, 2 place Le Verrier, 13248 Marseille
  cedex 04, France}
\email{lia@oamp.fr}

\begin{abstract}
The halo plays a crucial role in the evolution of barred galaxies. Its
near-resonant material absorbs angular momentum emitted from some of the disc
particles and helps the bar become stronger. As a result, a bar (oval)
forms in the inner parts of the halo of strongly barred disc
galaxies. It is thinner in the inner parts 
(but still considerably fatter than the disc bar) and tends to spherical at
larger radii. Its length increases with time, while always staying
shorter than the disc bar. It is roughly aligned with the disc bar,
which it trails only slightly, and it turns with roughly the same pattern
speed. The bi-symmetric component of the halo density continues well
outside the halo bar, where it clearly trails behind the disc bar. The
length and strength of the disc and halo bars correlate; the former
being always much stronger than the latter. If the halo is composed of
weakly interacting massive particles, then the formation of the halo
bar, by redistributing the matter in the halo and changing its shape,
could influence the expected annihilation signal. This is indeed found
to be the case if the halo has a core, but not if it has a steep
cusp. The formation and evolution 
of the bar strongly affect the halo orbits. A fraction of them becomes
near-resonant, similar to the disc near-resonant orbits at the same
resonance, while another fraction becomes chaotic. Finally, a
massive and responsive 
halo makes it harder for a central mass concentration to destroy the
disc bar.  
\end{abstract}


\section{The role of the halo in the evolution of barred galaxies}
\label{sec:role}

As shown by $N$-body simulations, barred galaxies undergo considerable
secular evolution (e.g. Hernquist \& Weinberg 1992; Debattista \&
Sellwood 2000; Athanassoula \& 
Misiriotis 2002 [AM02]; Athanassoula 2002 [A02];
Athanassoula 2003 [A03]; O'Neil \& Dubinski 2003;
Valenzuela \& Klypin 2003; Martinez-Valpuesta \& Shlosman 2004;
Athanassoula 2005a; Martinez-Valpuesta, Shlosman \& Heller 2005). 
The halo plays a major role in this, since it participates in the
angular momentum 
exchange within the galaxy. In an isolated disc galaxy with little or
no gas, angular momentum is emitted mainly by 
near-resonant material in the inner disc (bar region) and absorbed by
near-resonant material in the outer disc and in the halo (A02; A03). 
Since the bar is a negative angular momentum feature (Kalnajs 1971;
Lynden-Bell \& Kalnajs 1972), by loosing angular momentum it will grow
stronger (A02). 

\begin{figure} 
\begin{center}
\rotatebox{-90}{\includegraphics[width=10.5pc]{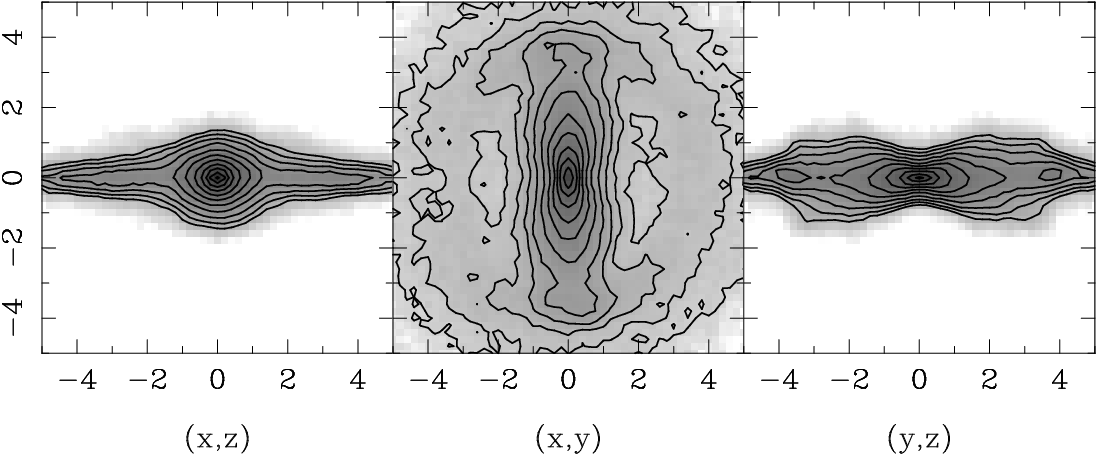}}
\vskip 8pt 
\rotatebox{-90}{\includegraphics[width=10.5pc]{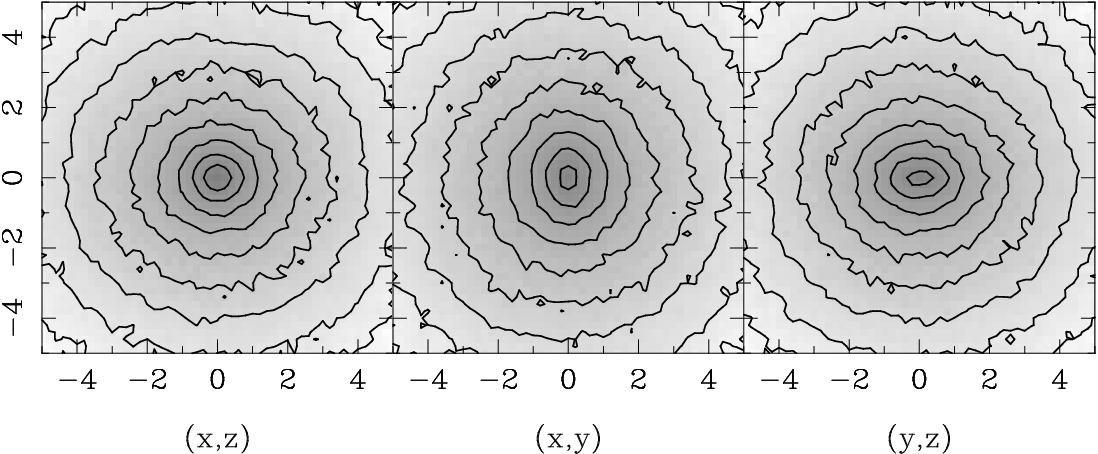}}
\end{center}
\vskip -3pt 
\caption[]{ Three orthogonal views of the disc (upper panels) and halo
component (lower panels). The central panels are face-on views, while
the others are edge-on; side-on for the right panels and end-on for
the left ones. Note that the halo does not stay axisymmetric, 
but forms an oval in its inner parts, which I call the
halo bar. }
\label{fig:halo-disc}
\end{figure}

\begin{figure}
\begin{center}
\rotatebox{-90}{\includegraphics[width=11.pc]{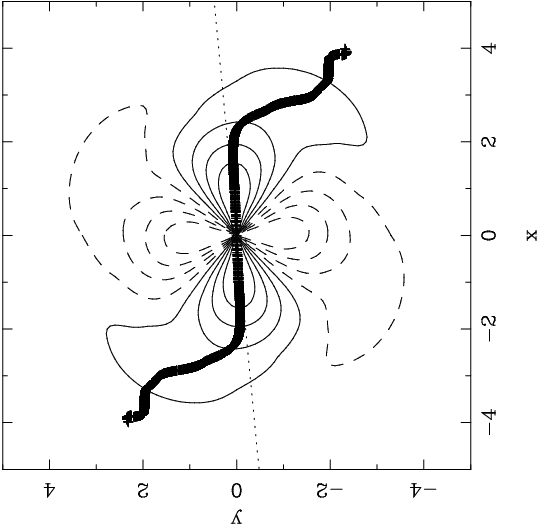}}
\rotatebox{-90}{\includegraphics[width=11.pc]{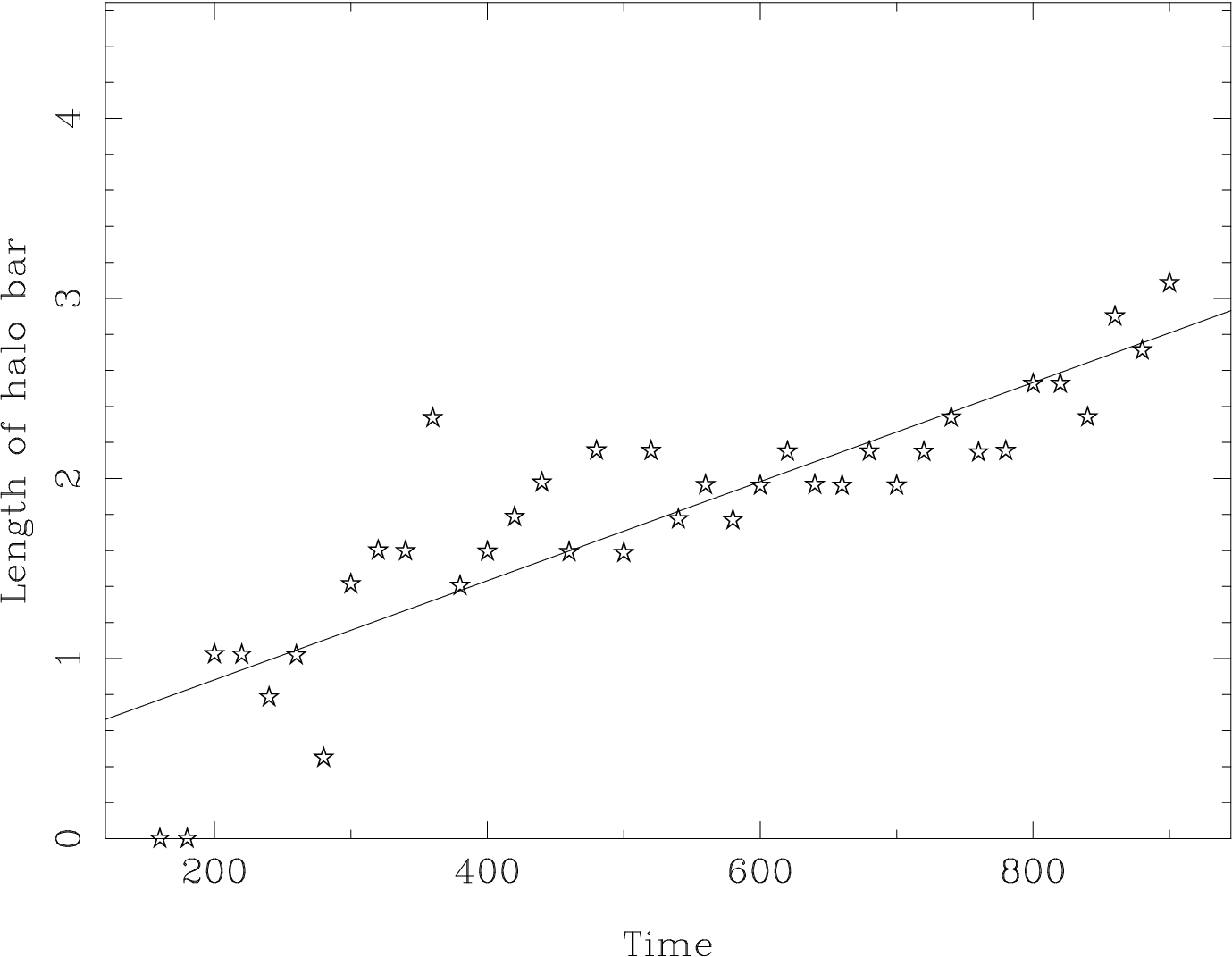}}
\end{center}
{\caption{{\bf Left panel} : Isocontours of the $l$=2 $m$=2 component of the
    halo mass 
  distribution on the galaxy equatorial plane. Positive isocontours are given
  with solid lines and negative ones with dashed lines. The thin
  dotted line gives the position angle of the disc bar and the thick
  line shows the phase of the halo bar. The latter is only plotted in
  regions where the amplitude of the $l$=2 $m$=2 component is at least
  equal to 4\% of its maximum amplitude. {\bf Right panel} : Length of
    the halo bar as a function of time. The solid line is a least
    squares fit, to guide the eye.}}
\label{fig:phase}
\end{figure}

Both for the disc and for the halo, there is more angular momentum gained
(or lost) at a given resonance if the density is higher there and if
the near-resonant material is colder. So, for equal amounts of mass,
the outer disc will absorb more angular momentum than the halo, because
it is colder. There is, however, considerably less mass in the outer
disc than in the halo, so the role of the halo can be very important,
or even predominant (A03). Thus, bars immersed in massive responsive
haloes can grow stronger than bars immersed in weaker haloes
(AM02; A03) and very
much stronger than bars immersed in rigid haloes (A02), since the
latter can not, by their formation, absorb any angular momentum. 

Since the halo plays such a crucial role in the evolution, it is
reasonable to expect that its properties will evolve with time, as do
the bar properties. This is indeed the case. Here, I
will particularly 
discuss the formation of an oval in the inner parts of the halo.
 
\section{The halo bar}
\label{sec:halobar}

Figure~\ref{fig:halo-disc}, from a simulation similar to those
described in AM02 or A03, shows three orthogonal views of the disc
(upper panels) and the halo (lower panels) from a simulation at a time well
after the bar has formed, i.e. while it is secularly evolving. The disc is
strongly barred, and seen side-on (i.e. edge-on, with the line of sight
along the bar minor axis) it shows a strong peanut. The evolution of
the halo, although less spectacular, is still quite clear. The inner
parts have lost their initial spherical symmetry and show a clear
prolate deformation, whose principal axes are roughly aligned with
those of the disc bar. Such a deformation is seen in all the
simulations with a strong bar that I analysed and has also been
discussed in Athanassoula (2005b [A05b]) and in Colin, Valenzuela \& Klypin
(2005). The axial ratio in the 
inner parts is roughly 0.7 and increases with radius. 

Analysing the halo mass distribution in spherical harmonics, it is
possible to get more information on 
the halo bar. The left panel of Figure~\ref{fig:phase} shows 
isocontours of the $l$=2 $m$=2 component,
i.e. of the component that best characterises the halo bar. In the
inner parts, the phase of this component seems roughly the same as the
phase of the disc bar (dotted line in the figure). In fact, it trails
it very slightly, of the order of a couple of degrees so that it is not
easy to discern it on the plot. I found similar phase differences at
other times and in other simulations I analysed. This phase difference
increases substantially at larger radii, so that the outer parts of the  
$l$=2 $m$=2 component look spiral-like. 
As can be seen in the right panel of Figure~\ref{fig:phase}, the length
of the halo bar increases substantially with time.

\section{Halo geometry and the WIMP annihilation signal}
\label{sec:WIMPs}

Athanassoula, Ling and Nezri (2005) studied the impact of halo shape
and geometry on the expected WIMP (weakly interacting massive
particle) annihilation signal from the galactic center. They find that
the asphericity has a strong impact on the annihilation signal when
the halo density profile has a core near the center, such as advocated by
observations (Bosma 2004 and references therein), but becomes less
significant for cuspy profiles, such as advocated by simulations
(e.g. Navarro, Frenk \& White 1997; Moore \tal 1999; Diemand \tal
2005), and negligible in the presence of a central spike.
 
\section{Orbital structure in haloes}
\label{sec:orbits}

As already mentioned in section~\ref{sec:role}, the halo contains a fair
fraction of near-resonant orbits. These are mainly 
trapped around Lagrangian periodic
orbits (see Fig 12 of A05b), or around
x$_1$-tree\footnote{The x$_1$ orbits are periodic orbits elongated
  along the bar and closing after
  one rotation and two radial oscillations (Contopoulos \&
  Grosb$\o$l 1989). The x$_1$-tree is the 3D extension of the
  x$_1$ family (Skokos, Patsis \& Athanassoula 2002) and comprises,
  except for the x$_1$ family, other 2D and 3D families bifurcating
  from x$_1$.}
periodic orbits (Fig 11 of A05b).

Halo particles which are near-resonant at a given time are not
randomly chosen from the initial halo distribution
function. Particles at near-inner Lindblad resonance had
initially preferentially smaller 
cylindrical and spherical radii. They also had preferentially smaller
values of $L_z$, the $z$ component of the angular momentum. Particles at
near-corotation had preferentially initially intermediate cylindrical and
spherical radii (i.e. they were not located initially in the innermost
or in the outermost regions). They had initially preferentially smaller
values of $|u_z|$, the $z$ velocity component, and larger values of
$L_z$ than average. To summarise, we can say that the halo particles
that will become near-resonant had initially properties similar to those
of the disc particles.

An important property for the evolution of the halo, and in general of
the galaxy, is the fraction of chaotic orbits in its population. The
most straightforward way of measuring this in simulations is to
calculate the complexity of each orbit, as introduced by Kandrup,
Eckstein \& Bradley (1997). As shown by these authors, the complexity
correlates well with the short term Lyapounov exponents, often used to
measure chaos. Of course, the exact value of the fraction of chaotic
orbits depends on where one sets the dividing line between a regular and a
chaotic orbit. A05b calculated this fraction for
different such threshold values and showed that it increases with the
strength of the disc bar and that in a strongly barred
galaxy as much as a quarter of the halo orbits can be chaotic.

\section{Black holes and central mass concentrations in barred galaxies
with live haloes}
\label{sec:CMC}

As shown both by periodic orbit calculations (Hasan \& Norman 1990;
Hasan, Pfenniger \& Norman 1993) and by $N$-body
simulations (Shen \& Sellwood 2004 and references therein) a central
mass concentration (CMC) affects the stability of x$_1$ orbits, making
them unstable and incapable of sustaining the bar. Thus, provided the
CMC is sufficiently massive and/or sufficiently centrally
concentrated, it can  destroy the bar, or at least considerably lower
its amplitude. These simulations, however, use a rigid halo and thus 
are not fully-self-consistent. 

\begin{figure}
\begin{center}
\rotatebox{-90}{\includegraphics[width=12.5pc,angle=90.]{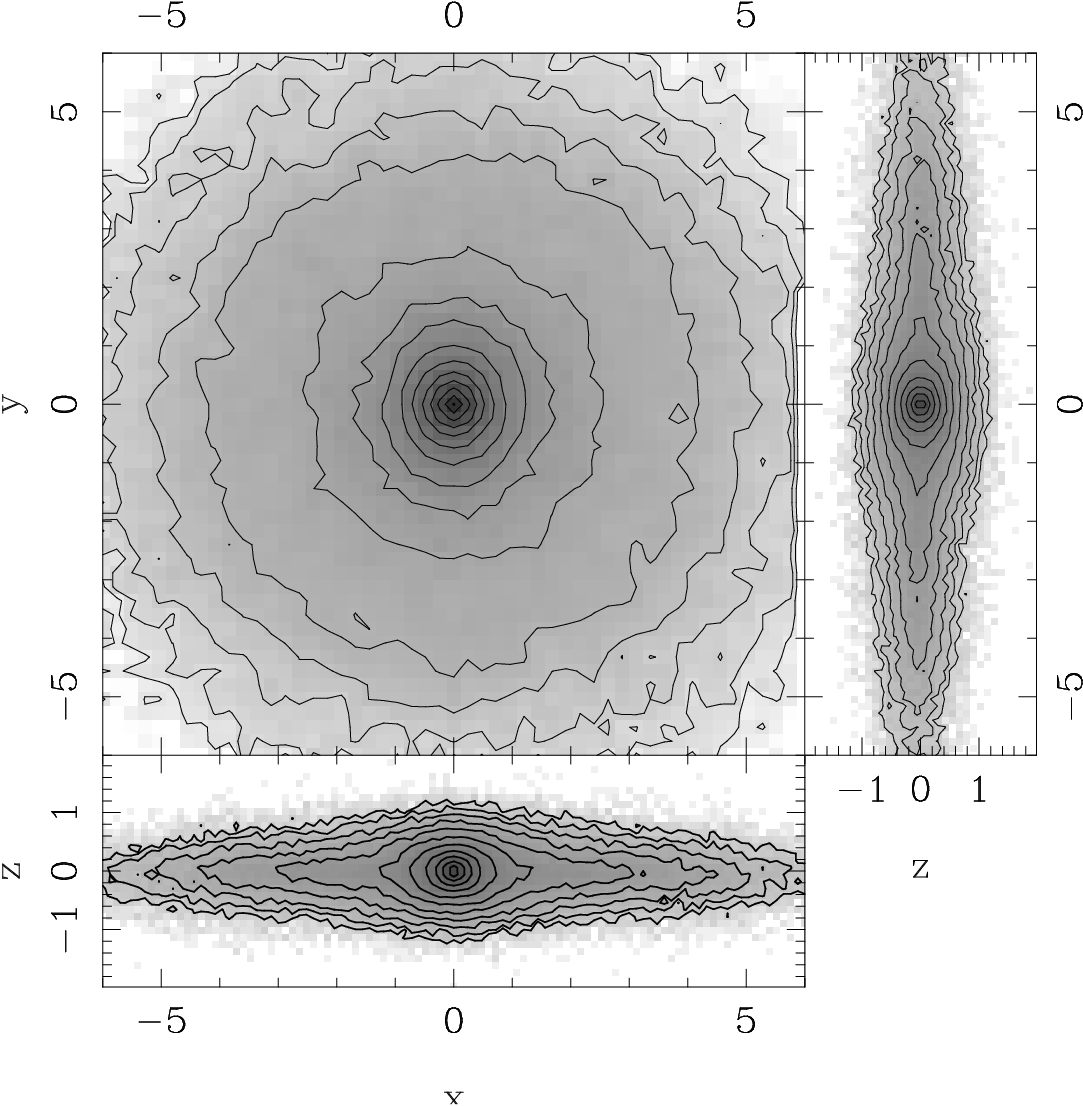}}
\rotatebox{-90}{\includegraphics[width=12.5pc,angle=90.]{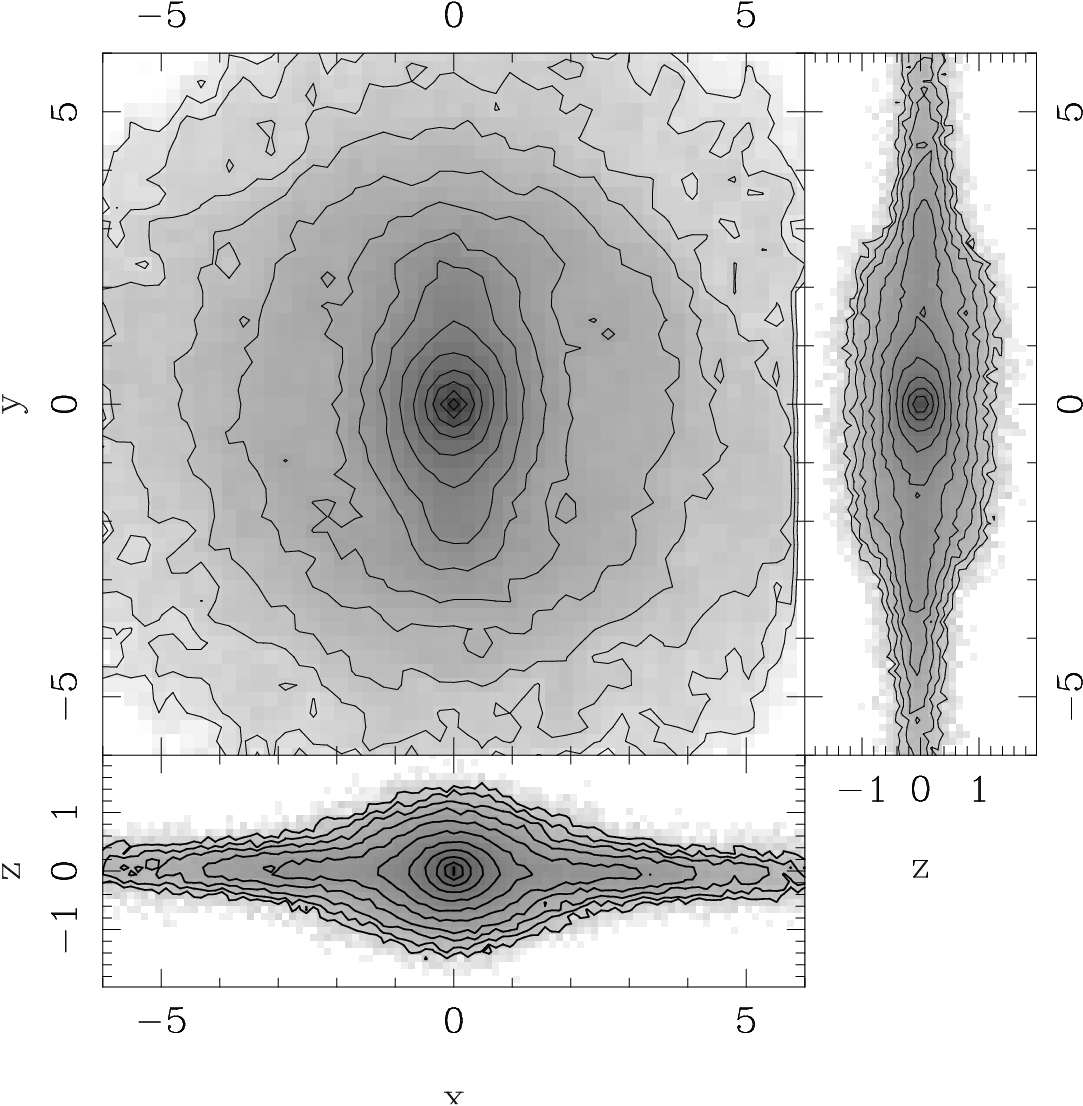}}
\end{center}
{\caption{Density distribution in the disc of a barred galaxy after
    the introduction of a CMC {\bf Left panel} :  MD-type model.
    {\bf Right panel} : MH-type model.}}
\label{fig:bhg_basic}
\end{figure}

Seen the important role that the response of the halo can play on the
evolution of barred galaxies, Athanassoula, Lambert \& Dehnen (2005) 
revisited this problem, now using a live halo, and I will here
summarise some of their results. They find that the effect of the CMC
depends drastically on  
the model. This is also illustrated in Figure~\ref{fig:bhg_basic}, which
compares the effect of a given CMC 
on two different barred galaxy models. The difference is quite
important. In the model on the left (MD-type, according to the
definition of AM02) the CMC totally destroyed the bar. But in the
model on the right (MH-type) the same CMC only decreased the 
bar strength, albeit substantially. In fact, the lowering of the bar
strength is mainly due to a decrease of the bar  
length and to more axisymmetric innermost parts. The latter can be
easily understood since this is the vicinity of the CMC. The CMC also causes
an increase of the bar pattern speed in all cases.   

The difference between these two models could be due to the
role of the halo in the two cases. Indeed, in MH-type models the
inner resonances in the halo are more populated, so that the halo can absorb
more angular momentum, compared to MD-type haloes (A03). This extra
angular momentum is taken from the bar and will, as discussed in
section~\ref{sec:role}, tend to increase its strength. It will thus
work against the CMC, whose effect will thus be lessened. This
is indeed what the simulations of Athanassoula et al. (2005) show.



\bibliographystyle{kapalike}


\end{document}